# Dynamic Tactile Sensing System and Soft Actor Critic Reinforcement Learning for Inclusion Characterization

John Bannan, Nazia Rahman, and Chang-Hee Won

*Abstract*— This paper presents the Dynamic Tactile Sensing System that utilizes robotic tactile sensing in conjunction with reinforcement learning to locate and characterize embedded inclusions. A dual arm robot is integrated with an optical Tactile Imaging Sensor that utilizes the Soft Actor Critic Algorithm to acquire tactile data based on a pixel intensity reward. A Dynamic Interrogation procedure for tactile exploration is developed that enables the robot to first localize inclusion and refine their positions for precise imaging. Experimental validation conducted on Polydimethylsiloxane phantoms demonstrates that the robot using the Tactile Soft Actor Critic Model was able to achieve size estimation errors of 2.61% and 5.29% for soft and hard inclusions compared to 7.84% and 6.87% for expert human operators. Results also show that Dynamic Tactile Sensing System was able to locate embedded inclusions and autonomously determine their mechanical properties, useful in applications such as breast tumor characterization.

*Index Terms*— Dynamic Tactile Sensing, Soft Actor Critic, Robotic Reinforcement Learning, Breast Tumor

## I. INTRODUCTION

Tactile sensing estimates mechanical properties such as size and elasticity of an object [1]. *Dynamic* tactile sensing extends this capability by capturing tactile information during motion for the perception of high spatial and temporal frequencies [2]. In contrast to static sensing, the dynamic motion of tactile sensors enhances sensitivity to reveal mechanical features that may otherwise remain undetected [2]. Dynamic tactile sensing has been applied across a diverse set of sensing tasks such as perception of liquid containers, fine surface features, and objects [2], [3], [4]. More recently, a Bayesian embedded object detection mapping method was reported in [5]. We extend this line of research by autonomously characterizing embedded inclusions. Our work aims to leverage dynamic tactile sensing for inclusion characterization with applications in breast tumor characterization.

Accurate breast tumor characterization is essential, as breast cancer is leading cause of death among women, and early detection can improve patient outlook [6], [7]. However, traditional screening modalities such as an MRI or mammography require specialized hospital settings and trained personnel, limiting their deployment in non-clinical environments. Moreover, the pain and discomfort associated with these methods are well documented and may lead to patients forgoing future examinations [8], [9]. A patient-friendly alternative we propose is the use of an autonomous robot to perform *dynamic* tactile sensing for breast tumor characterization. This would avoid MRI-related claustrophobia and reduce discomfort through precise force control compared to mammographic compression. Furthermore, surveys of automated robotic breast-exam systems have reported high patient acceptability and willingness to undergo robotic examination [10]. These considerations motivate the development of an autonomous robotic control framework that is capable of dynamic tactile sensing.

Dynamic tactile sensing in robots can be accomplished using reinforcement learning (RL), an artificial intelligence (AI) computational approach, where an agent learns by directly interacting with its environment [11], [12]. A key advantage of RL is that it does not require a predefined model of the environment, making it well suited for complex environments where a model is not known beforehand. RL algorithms maximize cumulative rewards, enabling agents to autonomously determine optimal actions through trial-and-error exploration [12]. Recent advances in RL have led to significant improvements in various applications such as gaming, robotic control, and computer vision [12], [13], [14].

Robotic *tactile* RL has been used in applications such as teaching a robot to flip through the pages of a book and ultrasound probe positing towards a target region [15], [16]. Other robotic systems have performed ultrasound and tactile characterization for vertebrae target localization through force

This work was supported in part by the National Science Foundation's Grant ECCS-2114675 and the Office of Vice President for Research, Temple University

(Corresponding author: John Bannan) John Bannan is with the Department of Electrical and Computer Engineering, Temple University, Philadelphia, PA 19122 USA (e-mail: john.bannan@temple.edu).

Nazia Rahman is with the Department of Electrical and Computer Engineering, Temple University, Philadelphia, PA 19122 USA (e-mail: rahman.nazia@temple.edu).

Chang-Hee Won is with the Department of Electrical and Computer Engineering, Temple University, Philadelphia, PA 19122 USA (e-mail: cwon@temple.edu).

feedback, though RL was not used for control [17]. The previously mentioned works use RL for trajectory optimization; however, they do not use RL for tactile imaging of inclusions or characterizing the target's mechanical properties. Previous work in our lab on dynamic sensing focused on inclusion size and depth estimation, using diffuse optical methods [18], [19]. However, our prior dynamic sensing did not use RL and was not applied to tactile sensing. Building on our previous work, we extend the dynamic sensing method by integrating RL and tactile sensing to characterize the location and mechanical properties of inclusions.

For our RL application, we propose to use the soft actor critic (SAC) algorithm to train a dual-arm robot (Baxter) to perform tactile sensing autonomously. SAC has been shown to provide stable learning and reliable performance for robotic trajectory control and path planning [16], [20]. SAC also leverages the benefits of sample-efficient off-policy learning while improving stability and convergence through the use of a maximum entropy framework; which reduces real time training and wear and tear on hardware [21]. This makes it suitable for robotic hardware implementation, where training time and mechanical stress warrant consideration. This is relevant for the use of tactile sensors, where significant repeated use may cause damage to the sensor. Thus, in this paper we introduce our Dynamic Tactile Sensing System (DTSS) that autonomously locates and estimates the mechanical properties of a target inclusion in a given region of interest (ROI) using a robotic arm and a trained tactile SAC model.

## II. DYNAMIC TACTILE SENSING USING SAC RL

This section introduces our DTSS that enables dynamic and autonomous interrogation of a ROI to locate and characterize embedded inclusions. The DTSS hardware consists of three components: a robot, the Tactile Imaging Sensor (TIS) (a tactile sensing device used in our lab's previous works [1], [18], [22]), and a laptop. The DTSS software consists of mechanical property computation software, robot firmware for control, and the algorithms developed and used in this paper.

The following subsections outline the principle of our TIS, the SAC-based control formulation, and the training and testing framework for the DTSS. We begin by describing the TIS, which forms the foundation of our proposed DTSS.

### A. Tactile Imaging Sensor

The TIS uses an optical method to collect tactile images which are converted into mechanical properties [1]. The TIS operates based on the principle of total internal reflection of light in a soft transparent Polydimethylsiloxane (PDMS) sensing probe. When an object deforms the probe, the light is scattered and captured by a CMOS camera (UI-3240CP-NIR, IDS Imaging Development Systems GmbH, Obersulm, Germany), creating a tactile image of the object [18], [22]. A force sensor (TE FX29, TE Connectivity, USA), with a range of 0 to 50 N, simultaneously records the applied normal force for each tactile image. Prior work [17], [21] relied on a trained human operator to apply a gradually increasing, *vertical*, normal force to obtain accurate results. Our DTSS uses RL and a robot to automate this process.

In RL, the main objective is to determine a policy which maximizes the expected sum of rewards [12], [22]. Among the many RL methods used for robotics, we adopt the SAC for its off-policy experience reuse, robustness, and strong sample efficiency in continuous space. This reduces training time and mechanical wear during learning [21], [23]. Therefore, the SAC algorithm is well suited for our DTSS. As required by SAC, we next define the state, action space, and reward for our system.

### B. Soft Actor Critic: State, Action Space, and Reward

The objective of our RL training is to teach a robot to take tactile data of unseen objects by using the SAC algorithm to maximize the reward derived from the tactile images. Thus, the robot's states, actions, and rewards must be defined before we can implement it. The robot's state is defined as the position of the end effector in robot's base frame and the action space consists of two movements in the Z-direction, either up or down by a fixed step. We only use the Z-direction as the TIS needs to have steady applied force in a vertically normal direction to operate properly. The reward is the sum of pixel intensity values in each tactile image, which increases when the probe deforms over an inclusion, thereby reinforcing the sensing behavior [1], [18], [22]. With the states, action space, and reward defined, we can now detail our training framework to utilize SAC. In order to implement this SAC algorithm on a robot, we need to train and test the Tactile SAC Model.

### C. Training of Tactile SAC Model

To implement the SAC algorithm on a robot, the following set up for training was developed as shown in Fig. 1. The Tactile SAC Model training framework comprises the robot, the TIS, and a laptop. The laptop handles communication and training of the DTSS through three main scripts: the main server script, the training script, and the environment script. Here, the TIS continuously transmits tactile data (the applied force and tactile images) to the main server script. In addition, the robot provides the current end effector position associated with each tactile image.

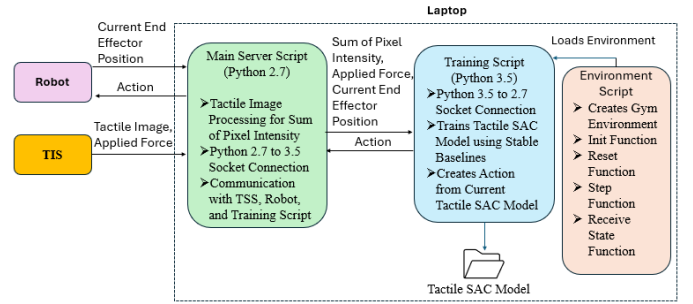

Fig. 1. Tactile SAC Model Training Framework

The main server script (Python 2.7) processes the tactile images to compute the sum of pixel intensities and communicates bidirectionally over a socket connection with the Training Script (Python 3.5). This socket connection bridges the limitations of our robot's software development kit requiring Python 2.7 and our desire to use SAC RL libraries in Python 3. The training script loads the environment from the environment script, implements the SAC RL using *Stable Baselines,* and generates robot actions based on the learned policy [24], [25]. The environment script defines the custom *Gym* environment, including initialization, reset, step, and state retrieval functions, where the reward corresponds to the sum of pixel intensity and the state is the end effector position [25]. Upon completion of training, the Tactile SAC Model is saved in the training script's file directory. To utilize our Tactile SAC Model on our robot, we test our model using the following testing setup.

### D. Testing of Tactile SAC Model

The testing setup, illustrated in Fig. 2, consists of the same robot, TIS, and laptop used in the training setup. The laptop communicates with the system and tests the model using four scripts: the main server script, the testing script, the environment script, and the camera server script. The main and environment scripts operate identically to those in the training framework. The testing script initializes the end effector position by a manually entered starting position and then queries the Tactile SAC Model from the training framework to generate an action. This action is transmitted to the main server, which sends the corresponding command to the robot. The robot executes the action and returns its updated current end effector position to the main script which relays it back to the testing script for further control cycles. During operation, the testing script sends the current end effector position and applied force to the camera server script and the TIS begins recording tactile data when the applied force is within a predetermined force range. The TIS then saves all tactile images to the laptop for post processing and analysis.

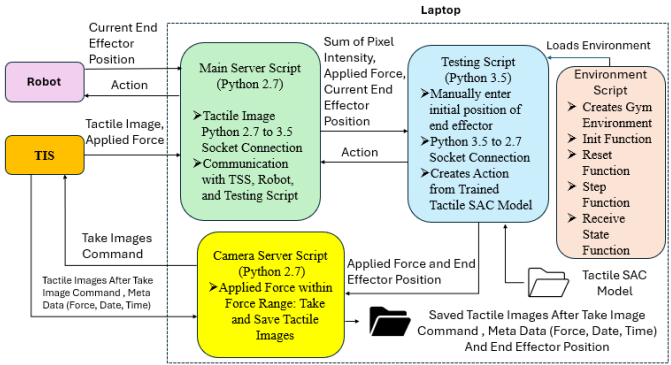

Fig. 2. Tactile SAC Model Testing Framework

The SAC RL testing framework provides the infrastructure for model execution and tactile data collection using the TIS with our robotic arm given a manual initial position. With a Tactile SAC Model, we will be able to dynamically sense inclusions using a robot manipulator. However, we must also be able to locate the inclusions without prior knowledge to utilize our Tactile SAC model. Therefore, in the next section, we introduce our Dynamic Interrogation algorithm that locates inclusions for a given region.

## III. DYNAMIC INTERROGATION TO FIND TARGET LOCATION

To determine the mechanical properties of inclusions, we first need to determine their location and quantity. Thus, we employ tactile exploration to extract relevant information of unknown objects through touch sensation [11]. This can be accomplished through tactile exploration procedures such as planned lateral movements and contour following [11], [26]. Thus, we introduce our method of Dynamic Interrogation as an exploration procedure to determine inclusion locations. This method utilizes dynamic tactile sensing with our dual arm robot alongside the DTSS and is split into two phases: Coarse Interrogation (CI) and Fine Interrogation (FI).

### A. Coarse Interrogation (CI)

The CI's objective is to collect tactile data over a defined ROI and return the location and number of possible inclusions. The CI algorithm visits specified locations in the ROI, illustrated as the "x" marked locations in Fig. 3, for our application example. Before we conduct CI, we must first define a ROI that we wish to interrogate. To define the ROI, we define a two-dimensional rectangular region with dimensions, $R_x$ and $R_y$, in mm with a coordinate origin, $(x_0, y_0, z_0)$, as seen in the bottom right corner in Fig. 3. Once the ROI has been determined we can select the starting point, $(x_s, y_s, z_s)$, where the CI will begin. In our specific application, the ROI is contained within a glass dish; therefore, to avoid possible collisions between the TIS and ROI boundary, we set the starting position well within the ROI. However, in other sensing settings, the starting position may be considered the same as the ROI origin.

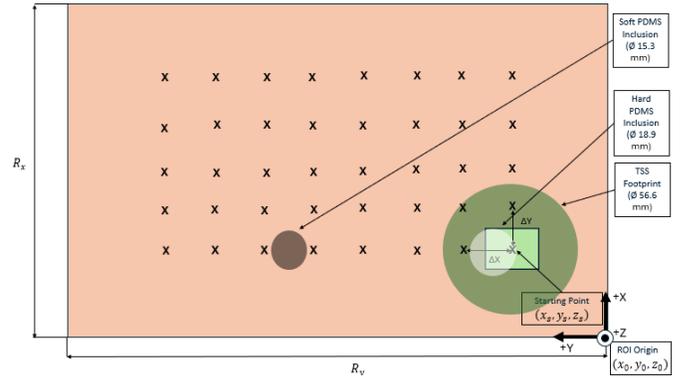

Fig. 3. Coarse Interrogation Application ROI

In Fig. 3. The "x" marked locations are evenly spaced in increments of ΔX and ΔY along the X and Y axes, respectively. These increments are determined by the footprint of the TIS and its sensing window. The CI algorithm begins by visiting the starting point and acquiring tactile data using TIS. The TIS then advances by ΔY to the next "x" location in the +Y direction and collects data. This is repeated until each location in the row has been visited, and data has been collected. Once every "x" has been visited in the row, the system increments by ΔX to the next row and repositions itself to the rightmost location in that row. The row scanning process is repeated for that row. This process is repeated until all target locations in the ROI are visited and imaged.

After the data is collected from each imaging location, we apply post-processing to remove noise and determine images that contain possible inclusions using region detection functions. We apply a predetermined diameter threshold and any region larger than this is treated as a potential inclusion. The ROI coordinate locations and associated robot end effector locations of these possible inclusion locations are then saved to be further investigated by the fine interrogation algorithm.

## B. Fine Interrogation (FI)

In the FI stage, the objective is to refine the locations of the inclusions found by the CI algorithm to more precisely determine the position and number of inclusions. The FI algorithm sequentially visits each location identified in the CI using the TIS and captures a tactile image under a constant specified applied force. For each image, the pixel centroid of the largest region is identified and the pixel offset between this centroid and the pixel image center along the X and Y direction is calculated. The TIS is then retracted and centered to this new centroid, and the process is repeated until the pixel offset between the tactile image centroid and image center falls below a predefined threshold. This final position is recorded as the estimated inclusion center.

After all CI locations have been processed through the FI, we analyze the spatial correspondence between them. We consider inclusions separated by less than a predefined distance threshold to be treated as the same inclusion. If the distance is greater than this threshold between two inclusions, they are treated as distinct inclusions. After we have identified all the unique inclusion centers, we apply our SAC tactile sensing model to image each location for detailed tactile imaging.

## IV. SYSTEM APPLICATION: BREAST TUMOR CHARACTERIZATION

Our application in this paper is motivated by breast tumor characterization, specifically to locate and determine the mechanical properties of embedded tumors. Mechanical property estimation is important for tumor characterization as properties, such as size and stiffness, serve as biomarkers for breast tumor pre-screening. Cancerous tissues, in particular, exhibit higher stiffness than benign tissue [22]. Previously in our lab, we developed a classification metric called the Risk Score that considers mechanical properties such as stiffness and size to classify benign and malignant [22]. The Risk Score is computed through analysis of the tactile images generated by the DTSS.

For our application, our DTSS is applied to dynamically sense and characterize embedded inclusions within PDMS phantoms that mimic the mechanical properties of breast tissue. The inclusions embedded in the phantoms are also made from PDMS and mimic the mechanical properties of tumors. The robot we use in our application is Baxter, a collaborative robot (cobot) specifically used for use with human interactions. Baxter uses compliant joints for safe interaction with human operators thus making it suitable for our application [27]. To utilize DTSS, we first trained our Tactile SAC model and validated it against tactile imaging done by human operators. From these tactile images, we estimate the mechanical properties of the inclusion.

### A. Mechanical Property Estimation

We estimate the size and deformation index as the mechanical properties, which are used to determine a Risk Score. This is important for tumor characterization application as malignant and benign tumors exhibit differences in mechanical properties, including size and stiffness [22].

#### 1) Tactile Size Estimation

We estimate the inclusion size from the tactile data captured by our TIS. This data is preprocessed to filter out noise by applying a median filter [22]. These images are further processed to estimate the inclusion size, $D$, following the method described in [22]. This method uses a 3D interpolation model utilizing the sum of pixel intensity, $I_p$, of each image and the associated applied force, $F$, and the modeled surface, $p_{i,j}$, to determine the size as: $D(F, I_p) = \sum_{i=0}^{i=n} \sum_{j=0}^{j=m} p_{i,j} F^i I_p^j$. Here, we modeled a third-order polynomial surface with indices $n = 2$ and $m = 1$.

#### 2) Tactile Deformation Index Estimation

Under the same size, depth, and applied force conditions, stiffer inclusions cause greater deformation of the TIS probe than the softer ones. This surface deformation is quantified by the Deformation Index (DI), defined as the slope of the graph plotted with sum of the change in pixel intensity versus the change in applied force defined as: $DI_i = \frac{\sum_{l=1}^{m} \sum_{k=1}^{n} \Delta I_i^{l,k}}{\Delta F_i}$, where $i = 1, 2, 3, \cdots$. The change in the sum of pixel intensity, $\Delta I_i$, and change in force, $\Delta F_i$, of every $i^{th}$ image is estimated with respect to a reference image [22].

#### 3) Tactile Risk Score Estimation

We estimate risk score, a unitless value, to classify an inclusion as more benign or malignant. This score is derived from the estimated $DI$ and the size, $D$, of the embedded inclusion, following the method described in [22]. In this study, the score ranges from 0 to 1, where 1 indicating an inclusion that is closer to malignant and 0 indicating that it is closer to benign. We define Risk Score as:

$$Risk\ Score = \left[\left(\frac{W_1 \times D}{D_{\max}} - \frac{W_2 \times DI}{DI_{\max}}\right)\right].$$

Where $W_1$ and $W_2$ are weights for the size and $DI$. $D_{\max}$ and $DI_{\max}$ are the maximum estimated size and maximum $DI$, respectively [22].

### B. SAC RL Validation Experiment

The Tactile SAC Model was trained using the DTSS which consisted of the TIS mounted to the end effector of our Baxter's left arm. A PDMS phantom with an embedded inclusion was placed underneath the TIS for imaging. The overview of the complete system and the PDMS phantom setup is given in Fig. 4. The Tactile SAC Model was trained following the training framework described in Section III.

To validate our Tactile SAC Model, we compared the data collected from the model to the tactile data collected manually by human operators under the same conditions. The manual data were collected by two operators who have expert level training and experience with the handling of the TIS.

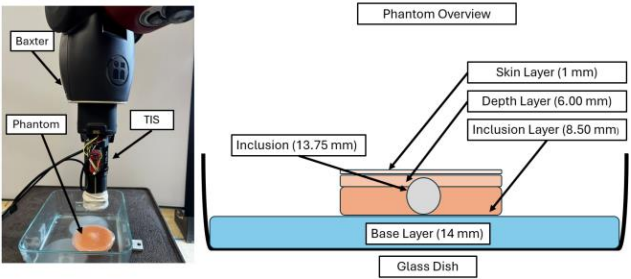

Fig. 4.  SAC Tactile Sensing Validation.

## C. Results of SAC Model Validation

We tested the Tactile SAC model on two inclusion cases: hard (18.9 mm) inclusion with a stiffness of 628 kPa and soft (15.3 mm) inclusion with a stiffness of 94.4 kPa using the setup in Fig. 4 over an applied force range of 0 to 10 N. We summarized the size, DI, and Risk Score estimation results for the soft and hard inclusions for the manually taken data and the Tactile SAC Model data in Tables I and II, respectively.

TABLE I. Size, DI, Risk Score Estimation (Manual Method)

| Human Operator Tactile Data | | | | | |
|---|---|---|---|---|---|
| | *True Size (mm)* | *Estimated Size (mm)* | *Size Error (%)* | *DI ($10^3$)* | *True Elasticity (kPa)* | *Risk Score* |
| **Soft** | 15.3 | 16.5 | 7.84% | 20.2 | 94.4 | 0.698 |
| **Hard** | 18.9 | 20.2 | 6.87% | 21.1 | 628 | 0.770 |

TABLE II. Size, DI, Risk Score Estimation (Tactile SAC Model)

| SAC Tactile Data | | | | | |
|---|---|---|---|---|---|
| | *True Size (mm)* | *Estimated Size (mm)* | *Size Error (%)* | *DI ($10^3$)* | *True Elasticity (kPa)* | *Risk Score* |
| **Soft** | 15.3 | 15.7 | 2.61% | 5.35 | 94.4 | 0.385 |
| **Hard** | 18.9 | 17.9 | 5.29% | 14.8 | 628 | 0.590 |

The size estimation error obtained with our proposed Tactile SAC Model is lower for both the hard and soft inclusions than that of the human operator, 2.61% vs. 7.84% for the soft inclusions and 5.29% vs. 7.84% for the hard inclusions. The *DI* ratio, defined as the ratio between the DI for the hard and soft cases, was computed as 1.04 for the manually acquired data and 2.77 for the Tactile SAC Model. Thus, the DI ratio for the Tactile SAC Model is larger, indicating a greater distinction between soft (benign) and hard (malignant) inclusions than the manually acquired data. The Risk Score for each method indicates that the harder inclusion (with higher kPa) is more malignant, as expected. However, the manually acquired data yields higher risk scores for both soft and hard inclusions compared to those produced by the Tactile SAC Model. The difference in DI and Risk Score between manual measurements and the Tactile SAC Model data is explained by the difference in how force was applied to the TIS. During Baxter data collection, force was exerted directly onto the top force sensor, while in manual data collection the TIS was held from the side, resulting in a different force distribution. Overall, the experimental results show that our proposed Tactile SAC Model produced better size estimation and distinction between soft and hard inclusions. Thus, our Tactile SAC model was deemed suitable for use in our DTSS, and we then applied our Dynamic Interrogation to an ROI with two unknown inclusions. We next describe the CI setup.

## D. Coarse Interrogation Setup

For our CI setup, we used a new rectangular PDMS phantom with the dimensions of $R_x = 165.1$ mm and $R_y = 215.9$ mm that contained two embedded inclusions. The phantom was opaque, made in a similar fashion to the one in Fig. 4 and was placed in a glass dish. However, the phantom had a larger inclusion layer (12 mm depth) than the one used in Fig. 4. In this case we embedded one hard inclusion and one soft inclusion, with the same size and stiffness as the ones used in our Tactile SAC Model validation experiment, in the inclusion layer. We defined the origin of the ROI in the Baxter base frame as: (0.6762, -0.1431, 0.1729). The starting location was defined in the ROI coordinate system as $(x_s, y_s, z_s)$ = (38.25, 38.25, 25). The true center of the hard and the soft inclusion were at (44.50, 51.50, -6) and (40.00, 118.50, -6), respectively, in the ROI coordinate system. Inclusions are indicated by the dark gray and white circles in Fig. 3.

## E. Coarse Interrogation Results

The results of the CI returned three locations with the following XY ROI coordinates: Location 1 (38.25, 53.25), Location 2 (38.25, 68.25), and Location 3 (38.25, 113.25). Our CI returned three inclusions, whereas the true number of inclusions is two. This is attributed to a single inclusion being present across multiple tactile images. Thus, with these three inclusions, we will interrogate each one using the FI algorithm.

## F. Fine Interrogation Setup

For the FI setup, we interrogate each of the three locations found by the CI and determine three corresponding more precise locations of the inclusions. The pixel offset selected threshold was set to 70 pixels, or 2.1 mm in distance for our DTSS camera set up. The resulting XY locations are as follows in ROI coordinate system: Location 1 (38.20, 62.00), Location 2 (44.00, 62.40), and Location 3 (44.00, 125.00).

The XY distances between Location 1 and Location 2 were found to be 5.8 mm, and thus below 6 mm so we consider this to be in the same location. This is because Baxter has a positional tolerance of ±5.00 mm, so we wish to have locations within 1.00 mm to be considered the same location, thus 6.00 mm. Thus, we selected Location 2 and Location 3 as the final locations of the inclusion, giving us two inclusions. It is important to note Location 2 was selected arbitrarily over Location 1, as they both considered the same point. The resulting distance between the true center of the hard and soft inclusion was calculated as 10.9 mm and 7.63 mm. The error in the resulting distance can be explained by the positional noise of the Baxter robot and the pixel threshold that was selected. Combining the 5.00 mm positional tolerance of the Baxter and the 2.10 mm positional threshold for the FI, we would have a total of around 7.10 mm of error resulting for the system itself. With the location and number of inclusions found, we then applied our Tactile SAC Model to image the possible inclusions.

### G. Fine Interrogation Results

For the two points considered, the Tactile SAC Model took tactile images, and the mechanical properties were estimated. The size, DI, and Risk Score for Target Locations 3 and 2 are found in Table III.

TABLE III. Size, DI, Risk Score (Fine Interrogation)

| | Target | True Size (mm) | Estimated Size (mm) | Size Error (%) | DI ($10^3$) | True Elasticity (kPa) | Risk Score |
|---|---|---|---|---|---|---|---|
| Soft | 3 | 15.3 | 15.7 | 2.61% | 3.78 | 94.4 | 0.335 |
| Hard | 2 | 18.9 | 16.5 | 12.6% | 9.65 | 628 | 0.532 |

Reviewing Table III., the size estimation error was slightly larger for the hard inclusion than for the soft inclusion. The increased error for the hard inclusion may be attributed to the increased depth of inclusion layer in the PDMS phantom. The DI ratio, however, was found to be 2.55, which is like DI ratio used in the Tactile SAC model validation experiment. The Risk Score also indicates that the hard inclusion is more malignant than the soft inclusion, as expected, and similarly is consistent with the results found in Table II.

## V. Conclusion

We developed a Dynamic Tactile Sensing System that autonomously locates and senses the mechanical properties of unknown inclusions using a robotic arm and a trained SAC RL model. A Dynamic Interrogation algorithm was developed that uses a coarse and fine interrogation to precisely determine the number and location of unknown embedded inclusions. A Tactile SAC model was also developed and trained to mimic human like tactile sensing using a dual arm robot with pixel intensity as a reward. Validation experimental results show that our proposed Dynamic Tactile Sensing System provides better size estimation when compared to manual human operation. We conclude that our proposed Tactile SAC model algorithm can effectively be applied to tactile sensing for inclusion characterization. Our current work shows that our Dynamic Tactile Sensing System is able to autonomously locate and estimate the mechanical properties of inclusions better than an human operator.

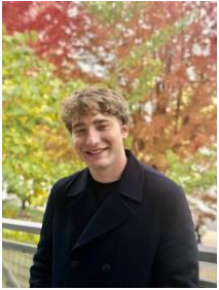

**John Bannan** received the B.Sc. degree in electrical engineering from the University of Notre Dame, Notre Dame, IN, USA, in 2022 and his Masters in Electrical Engineering in 2025 from Temple Univeristy, Philadelphia, Pa, USA. He is currently pursuing a Ph.D. degree in electrical engineering at Temple University. His current research interests are tactile sensing, robotics, control theory, dynamic sensing systems, and reinforcement learning.

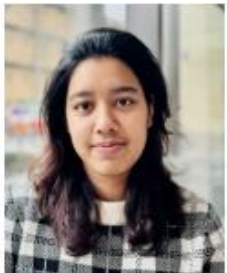

**Nazia Rahman** received the B.Sc. degree in electrical and electronic engineering from Chittagong University of Engineering and Technology, Chattogram, Bangladesh, in 2017. She is currently pursuing the Ph.D. degree in electrical engineering with Temple University, Philadelphia, PA, USA. Her research interests include tactile sensor, multispectral sensor, vibroacoustic system, reinforcement learning, and optimal control.

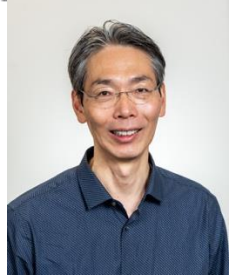

**Chang-Hee Won** received the Ph.D. degree in electrical engineering from the University of Notre Dame, Notre Dame, IN, USA, in 1995. He is a Professor at the Department of Electrical and Computer Engineering and the Director of Control, Sensor, Network, and Perception (CSNAP) Laboratory, Temple University, Philadelphia, PA, USA. Previous to coming to academia, he worked at the Electronics and Telecommunications Research Institute as a Senior Research Engineer. He is currently actively guiding various research projects funded with the National Science Foundation, the Pennsylvania Department of Health, and the Department of Defense. He published over 120 peer-reviewed articles and received multimillion dollars of research funding as a Principal Investigator from industry, state, and federal funding sources. His research interests include tactile sensing, optimal control theory, spectral imaging, reinforcement learning, and dynamic sensing systems.